\documentclass[final,5p,times,twocolumn]{elsarticle}

\usepackage{color,amssymb}

\usepackage{hyperref}

\usepackage{amsmath} 
\usepackage{multicol} 
\usepackage{graphicx} 
\usepackage{nameref}
\usepackage{cleveref}
\newcommand*{\fullref}[1]{\hyperref[{#1}]{\autoref*{#1} \nameref*{#1}}} 
\journal{Physics Letters B}

\begin{document}

\begin{frontmatter}


\title{
Implications of ALP-photon conversion for the diffuse gamma-ray background associated with high-energy neutrinos
}


\author[UC,INR]{Kirill Riabtsev}
\ead{kr548@cantab.ac.uk}


\affiliation[UC]{organization={Department of Applied Mathematics and Theoretical Physics, University of Cambridge},
            addressline={Wilberforce Road}, 
            postcode={CB3 0WA}, 
            state={Cambridge},
            country={UK}}

\affiliation[INR]{organization={Institute for Nuclear Research of the Russian Academy of Sciences},
            addressline={60th October Anniversary Prospect 7A}, 
            postcode={117312}, 
            state={Moscow},
            country={Russia}}

            
\date{\rm INR-TH-2023-017}
\begin{abstract}

Some fraction of the diffuse photon background is supposed to be linked to high-energy neutrinos by astrophysical mechanisms of production and electromagnetic cascades. This article presents a simulation study of axion-like particles (ALPs) implications for that component, exploiting transport equations. Alternations of that spectrum due to ALP-photon conversion in the intergalactic magnetic field (IGMF) in the cases of various ALP parameters and mixing regimes at sources are studied. The results indicate considerable influence of IGMF-conversion on the ALP-photon flux even in the case of inverse ALP-photon coupling constant $M$ equal to $10^{11.5}$ GeV and some residual effects in the case of $M=10^{12}$ GeV. Furthermore, the scenario shows another aspect of a complex multimessenger interplay between IceCube and Fermi data, to a certain extent relieving the tension between them, the existence of which is commonly interpreted to be a signature of limited transparency of the neutrino sources population.

\end{abstract}


\end{frontmatter}


\section{Introduction, motivation of the paper}
\label{sec:Intro}

Axion-like particles (ALPs, sometimes for simplicity called axions in our paper) are hypothetical particles that are predicted by extensions of the Standard Model and can interconvert with photons in an external magnetic field. This became the basis for most observational ALP searches.\cite{Mastrototaro_2022, Troitsky_2022, Ajello_2016, Galanti:2022iwb, DeAngelis:2011id, Marsh_2017, Liang_2021, Irastorza_2018, Guo_2021, meyer2016searches, Troitsky:2023uwu, Pant:2023khq}. At the same time, two main mechanisms have been discussed: the first involves conversion near the source, propagation through intergalactic space, and later reconversion in the Milky Way (MW) magnetic field; the second involves conversion of photons and axions on their way from the source in the intergalactic magnetic field (IGMF), despite scarce information and constraints on them (for the latter Ref.~\cite{De_Angelis_2007,Mirizzi_2017,Galanti:2018myb, S_nchez_Conde_2009}, one of most recent Ref.~\cite{Kachelriess:2023fta}, also discussed in the context of supernova dimming in Ref.~\cite{Cs_ki_2002, M_rtsell_2002,Christensson_2003}). Apart from Gaussian fields, other structures of magnetic fields are considered in the literature, for example, domain-like structures and their variations (Ref.~\cite{Kachelriess:2023cjh,Galanti_2018,S_nchez_Conde_2009}), various computational approaches and programs are described (for instance, Ref.~\cite{Meyer_2021, Kachelriess:2021rzc}).
Though many of the listed works considered the joint influence of conversion and pair production process (PP), as far as we know none of them involved inverse Compton scattering (ICS). 
ALP-related effects outside of conversion have also been studied for relieving multimessenger tensions (for instance, in Ref \cite{Kalashev_2019}).


Electromagnetic cascade (later e-m cascade) is the process of sequential PP and ICS processes on background radiation, due to which the photons reach the observer at lower energies than emitted (for a review see Ref.~\cite{Berezinsky_2016}). Different approaches and programs (except for the one we are using in our studies) for the modeling of e-m cascades exist \cite{Blanco_2019,Batista_2016}.

The production of high energy gamma-rays in astrophysical environments is connected with the emission of neutrinos via so-called $pp$-mechanism (in other words, hadronic interactions, named that way because of the protons or nucleus colliding over nucleus at rest) and $p\gamma$-mechanisms (or, alternatively, photohadronic interactions, in which the target is replaced by gamma-rays), so neutrinos are supposed to have their injected photon counterpart with energies and fluxes of the same order of magnitude \cite{Palladino_2020}. In the context of this work it allows us to normalize the photon flux and compare that contribution with the diffuse background detected by Fermi-LAT, thus deepening our understanding of its nature and composition, which is a complex problem (see Ref.~\cite{Fornasa_2015}). The synergy of photons and neutrinos in the ALP context has also been studied, though for sub-PeV energies, in Ref.~\cite{Eckner_2022}.

We assume, for simplicity, that the full or at least most of IceCube flux has extragalactic origin (see discussions in Ref.~\cite{Murase_2016} and references therein). However, it is important to notice the recent studies, providing indications about the fraction of neutrinos with galactic origin \cite{Kovalev_2022, Albert_2023, 2023}.

We remain agnostic about the specifics and nature of individual sources. There are indications suggesting non-transparency of the sources, importantly the IceCube-Fermi spectral tension (see the report by the IceCube collaboration \cite{IceCube:2020wum} and Ref. \cite{Murase_2016, kistler2015problems}). Additionally, scenarios involving non-transparency of the sources are proposed to explain the lack of coincidences in the neutrino detections associated with specific sources, see Ref. \cite{Kun:2023uld}. In this work, we model the sources as gamma-ray transparent, thus avoiding arbitrariness and exploring how the introduction of axion-like particles (ALPs) could alleviate the tension between the IceCube and Fermi spectra, which otherwise supports the argument for non-transparency.

For our purposes we set out to use the approach of transport equations, adopt the code TransportCR \cite{kalashev2014simulations} as a tool for simulations. At this point, we primarily aim to catch the key features and possibilities for the IGMF-mixing on the upper bound ($\sim 1$~ nG) to modify the spectra rather than draw robust constraints on the parameter space.

We organize the paper as follows. First, we derive modified transport equations (Eq.~\ref{eq:Transport}), describing cascade development influenced by ALP-photon (later sometimes abbreviated as ALPh in this paper) conversion in Subsec.~\ref{subsec:ALP} and ~\ref{subsec:transport}. Next, we list the parameters that were included in the model and the motivation of their choice in Subsec.~\ref{subsec:parspace},~\ref{subsec:injection},~\ref{subsec:axinjection} and ~\ref{subsec:IGMF}. Finally, we present the results of our modeling (propagation of the ALP-photon flux from the sources until it reaches the MW magnetic field) and discuss their features in Sec.~\ref{sec:Results}.

\section{Mathematical formalization of the problem, parameters of the model}
\label{sec:main1}
\subsection{Basics of ALP-photon (ALPh) conversion}
\label{subsec:ALP}

ALPs are coupled to the electromagnetic field via the interaction term of the Lagrangian, see Ref.~\cite{Fairbairn_2011}: 

  \begin{equation}
    \mathcal{L}_{int}=-\frac{1}{4}\frac{a}{M}F_{\mu \nu}\widetilde{F}^{\mu \nu},
    \label{eq:Lagrangian}
\end{equation}

where $F_{\mu \nu}$ is the electromagnetic stress tensor and  $\widetilde{F}^{\mu \nu}$ is its dual, $a$ is the pseudo-scalar ALP field, $g=\frac{1}{M}$ denotes photon-axion coupling constant and $M$ is its inverse, $m$ further denotes the ALP mass.

As presented in Ref.~\cite{Fairbairn_2011} for constant magnetic fields and as holds true in the case of non-uniform magnetic fields for order-of-magnitude estimates, there are two main conditions for ALPh mixing to be significant (which is called the strong, or maximal, mixing). The first one is given as (for the simplest estimates, not including terms related to electron density, energy density of the magnetic field, and photon attenuation in the expression; see the definitions of $\Delta_{a \gamma x,y}$ and $\Delta_{a}$ below) 

\begin{equation}
   2 \Delta_{a \gamma x,y} \gtrsim  \Delta_{a}
   \label{eq:Strongmix}
\end{equation}

The second one is given by the relation between the characteristic length (that is, the size of the region in which mixing should be effectively carried out, it is for instance either the physical size of the source or attenuation length, that effectively may prohibit conversion of photons into axions due to attenuation on background light) and the oscillation length:

\begin{equation}
   {L_{osc}} \cong \frac{\pi}{\Delta_{osc}}\lesssim {L}
   \label{eq:Lengthmix}
\end{equation}

In these expressions $\Delta_{a \gamma x,y}$, $\Delta_{a}$ and $\Delta_{osc}$ are given by

\begin{equation}
    \Delta_{osc}=\sqrt{4\Delta_{a \gamma x,y}^2+\Delta_{a}^2},
    \label{eq:Deltaosc}
\end{equation}
\begin{equation}
    \Delta_{a \gamma x,y}=\frac{B_{x,y}}{2M}=153 \times 10^{-9} \ \left(\frac{B_{x,y}}{1 \ \mbox{nG}} \right) \left( \frac{10^{10} \ \mbox{GeV}}{M}\right) \  \mbox{pc}^{-1},
    \label{eq:Deltaagxy}
\end{equation}
\begin{equation}
    \Delta_{a}=\frac{m^2}{2E}= 7,8 \times 10^{-4} \left(\frac{m}{10^{-7} \ \mbox{eV}} \right)^2 \left( \frac{10^{12} \ \mbox{eV}}{E}\right) \  \mbox{pc}^{-1}, 
    \label{eq:Deltaa}
\end{equation}

where $B_{x,y}$ is the magnetic field strength in one of the directions traverse to the direction of propagation.  $E$ is the photon (axion) energy.


%

\subsection{Modified transport equations, employed in the calculations}
\label{subsec:transport}

We will treat ALP-photon mixing within the formalism of the density matrix, the evolution of which follows Liouville equation, for a review see Ref.\cite{vogel2017diffuse, Kartavtsev_2017}. 

Note that the expressions below are valid for cases where we can neglect terms associated with electron density and energy density of the magnetic field (see the expressions in the same references). This is applicable in intergalactic space, where the electron density is sufficiently low (see Ref. \cite{Kachelriess:2021rzc}), and the energy density of the magnetic field is low enough, with an upper bound as mentioned above.

Using the same mathematical technique, as suggested in Ref.~\cite{Zhang_2013} for the study of neutrino oscillations. and denoting $\rho_0$, $\rho_1$, $\rho_2$, $\rho_3$, $\rho_4$, $\rho_5$, $\rho_6$, $\rho_7$, $\rho_8$ as coefficient of expansion of  the density matrix in the basis of the Gell-Mann matrices together with the identity matrix (for a review on their properties, including (anti)commutation relations, see \cite{GellRef2}), we obtain: 

\begin{equation}
\begin{aligned}
    \begin{cases}
        \frac{dN_{ph1}}{dt}=2\Delta_{a \gamma x}\rho_5+...~, \\
        \frac{dN_{ph2}}{dt}=2\Delta_{a \gamma y}\rho_7+...~, \\
        \frac{dN_{a}}{dt}=-2(\Delta_{a \gamma x}\rho_5+\Delta_{a \gamma y}\rho_7), \\
        \frac{d\rho_1}{dt}=\Delta_{a \gamma x}\rho_7+\Delta_{a \gamma y}\rho_5 - \Gamma \rho_1, \\
        \frac{d\rho_2}{dt}=\Delta_{a \gamma x}\rho_6-\Delta_{a \gamma y}\rho_4 - \Gamma \rho_2, \\
        \frac{d\rho_4}{dt}=\Delta_{a \gamma y}\rho_2-\Delta_{a}\rho_5 - \frac{1}{2}\Gamma \rho_4, \\
        \frac{d\rho_6}{dt}=-\Delta_{a \gamma x}\rho_2-\Delta_{a}\rho_7 - \frac{1}{2}\Gamma \rho_6, \\
        \frac{d\rho_5}{dt}=-\Delta_{a \gamma y}\rho_1+\Delta_{a}\rho_4+\Delta_{a \gamma x}(N_{a}-N_{ph1}) - \frac{1}{2}\Gamma \rho_5, \\
        \frac{d\rho_7}{dt}=-\Delta_{a \gamma x}\rho_1+\Delta_{a}\rho_6+\Delta_{a \gamma y}(N_{a}-N_{ph2}) - \frac{1}{2}\Gamma \rho_7, \\
        \frac{N_{e}}{dt}=...
    \end{cases}
\end{aligned}
\label{eq:Transport}
\end{equation}
 where $\Gamma$ is the photon absorption rate and $N_{ph1}$, $N_{ph2}$ denote the number of photons of different polarizations and $N_{a}$ denotes the number of axions, after "..." terms of usual transport equations, that account only for ICS, PP and source density, are followed, $N_{e}$ denotes the number of electrons/positrons (as in "usual" transport equations describing e-m cascades (for a detailed description see Ref.~\cite{Lee_1998}). Note that $\rho_0$, $\rho_3$, $\rho_8$ are not included, since, as suggested in \cite{Zhang_2013}, we keep the diagonal terms in the natural basis $N_{ph1}$, $N_{ph2}, N_{a}$.

\subsection{ALP parameter space, the region selected for the simulations}
\label{subsec:parspace}
\begin{figure}
\centering
\includegraphics[width=\linewidth]{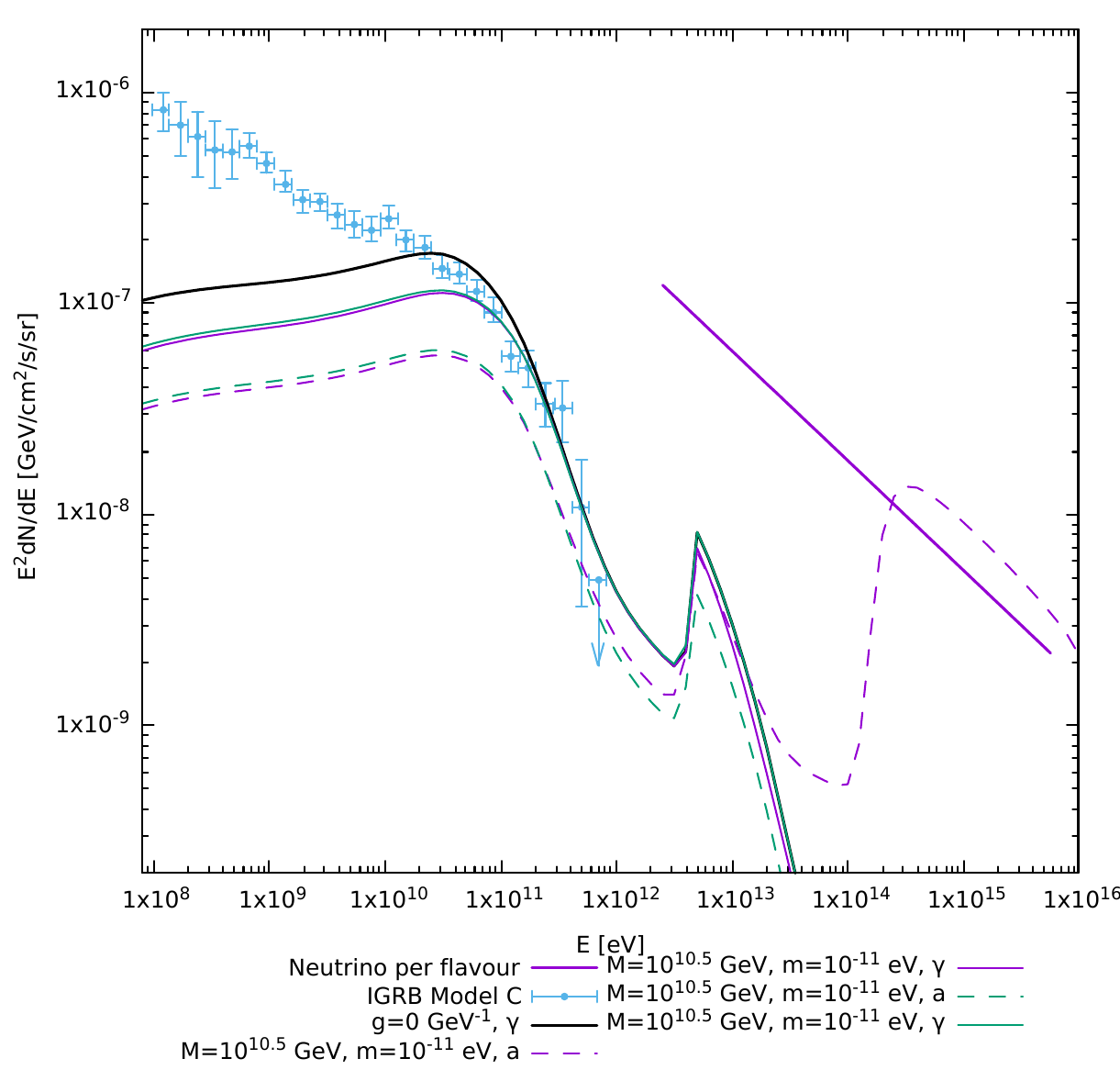}
\caption{\label{fig:multimessenger}
To \fullref{subsec:messenger}; green color here is used for the case of no mixing at the sources and purple color is used for the case of strong mixing at the sources. ALP parameters are specified in the legend (they are the same here for both cases involving axions); IceCube spectrum from analysis \cite{naab2023measurement}; Fermi spectrum from \cite{Ackermann_2015}; "$g=0 ~ \mbox{GeV}^{-1}, \gamma$" corresponds to the standard scenario, "$\gamma$"/"a" labels photons(the sum of both polarizations)/axions; solid/dotted lines are used for photons/axions here and further in \Cref{fig:evolution,fig:30to55,fig:axmixhigheren,fig:fermimixnoax,fig:fermimixwithax}).
}
\end{figure}

\begin{figure}
\centering
\includegraphics[width=\linewidth]{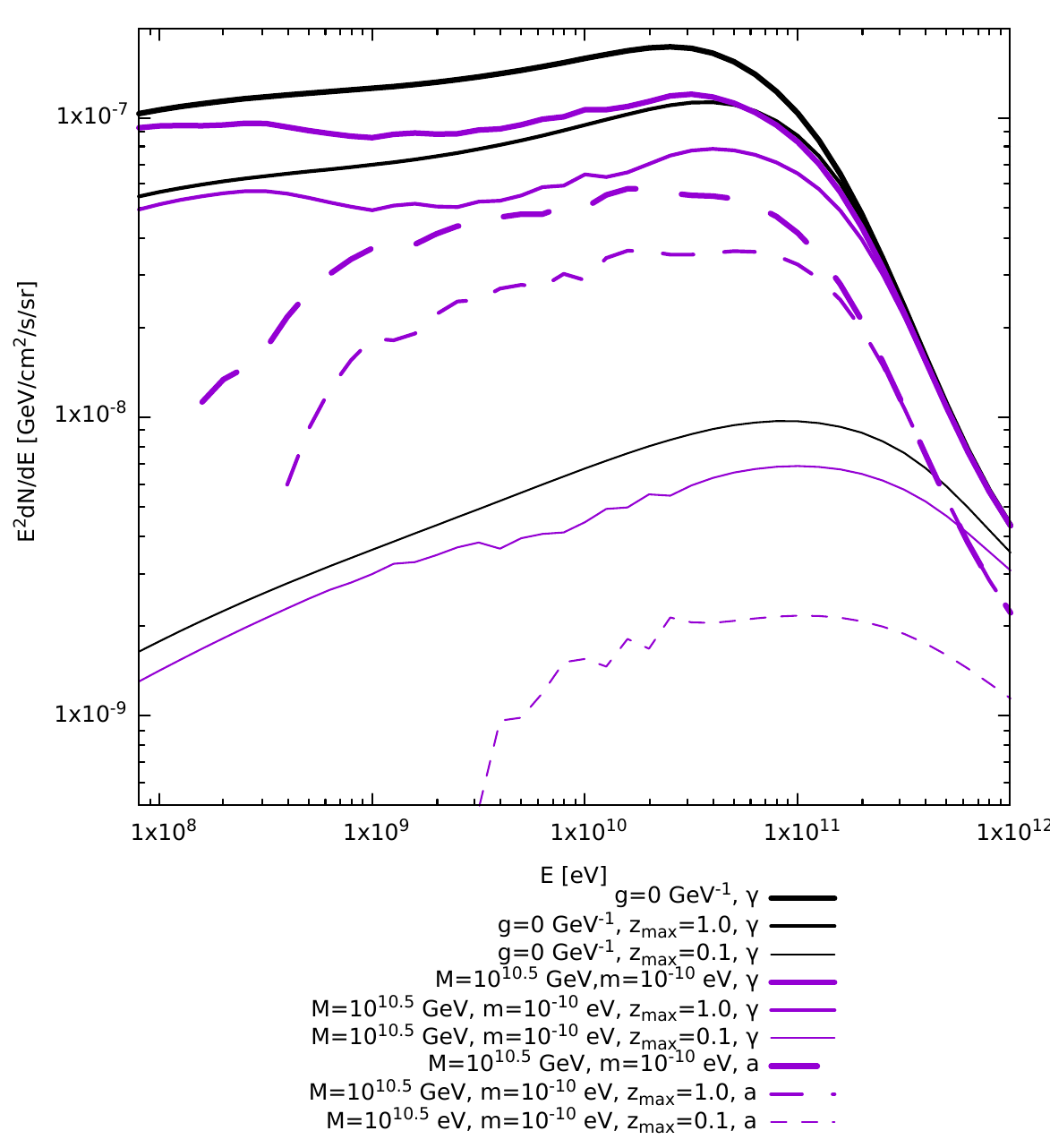}
\caption{\label{fig:evolution}
To \fullref{subsec:evolv}; the same ALP parameters everywhere (specified in the legend); the width of the lines is adjusted according to the maximal redshifts.
}
\end{figure}

We consider ALP inverse coupling up to $10^{12}$ GeV, which (almost) prohibits conversion due to the oscillation length exceeding attenuation and coherence (see Subsec.~\ref{subsec:IGMF}) lengths, and focus mostly on the region not constrained by CAST (see Ref.~\cite{ParticleDataGroup:2022pth} for a review on the current limits and their nature). Axion mass $m$ is taken according to the condition (Eq.~\ref{eq:Strongmix}) in order to "turn off" the effect of ALP-photon conversion below Fermi or IceCube ranges (see Subsec.~\ref{subsec:cases} and the legend of ~\Cref{fig:multimessenger,fig:evolution,fig:30to55,fig:axmixhigheren,fig:fermimixnoax,fig:fermimixwithax} therein , where the ALP parameters are specified).

\subsection{IGMF model}
\label{subsec:IGMF}

 We consider a divergence-free Gaussian turbulent magnetic field with a Kolmogorov spectrum, zero mean, and root mean square (RMS) $B_{0}$ equal to $1$ nG, generating it following the procedure described in Ref.~\cite{Meyer_2017}. Minimal and maximal spatial scales are picked to be equal to $10^{2} ~\mbox{Mpc}$ and $10^{-1} ~\mbox{Mpc}$, which corresponds to the coherence length of $20 ~\mbox{Mpc}$, according to Ref.~\cite{Blytt_2020}. Note that Ref.~\cite{Pshirkov_2016, Planck:2015zrl} give values that are close to our model, while Ref.~\cite{Neronov:2021xua} predicts more stringent limits. In the latter case the presented results of simulations would remain valid, but for stronger coupling, that is, to $M$ approximately half an order of magnitude lower than in  Subsec.~\ref{subsec:cases}.

 The evolution of the magnetic field as $B \propto B_{0} \ (1+z)^2$ was considered. \cite{DeAngelis:2011id, Grasso_2001}

\subsection{Other parameters}
\label{subsec:injection}

We use a single power law from Ref.~\cite{naab2023measurement} for the neutrino injection spectrum, not extrapolating it beyond the fit. We choose the redshift distribution from Ref.~\cite{Kachelrie__2017} for the detailed investigations of the results dependence on ALPs parameters.

Fig.~\ref{fig:comparisonredshifts} presents the comparison with some other redshift distribution models, taken from \cite{Hasinger:2005sb}(Table 6, plotted in Fig.9 therein) and specified on the plot according to the classification given therein, the pairs of photon spectra for axion and non-axion case are presented; ALPs parameters are the same as in Fig. \ref{fig:multimessenger}, see the discussion below. As can be seen, the choice of the redshift distribution affects the energy range around $10^{11}$ eV and above. The effect of photon flux reduction gets weaker for the distributions with the peak located at higher redshifts, and at some point, it flips with a slight enhancement, which nevertheless occurs for the distributions with less tension.

We take the relationship between photons and neutrinos injected as in the $p\gamma$-mechanism (see Ref.~\cite{Ahlers_2018}), which does not affect the order of magnitude of the results and the spectral shapes, while remaining agnostic about the nature of the sources, thus adjusting the injection spectra of photons, as we mentioned above in this section, to the observational fit and not discriminating between different shapes of more realistic spectra produced by $p\gamma$ and $pp$. 

For a detailed breakdown of the latter, see Ref. \cite{Murase_2016}. Note that the qualitative feature of the photon spectrum in the $pp$ case is the extrapolation of it to lower energies, which increases the disagreements between Fermi and IceCube data (as also noted in \cite{kistler2015problems}). Such an extrapolation would generally stretch the photon spectra to energy ranges beyond the registered neutrinos. At the same time, there are other indications in favour of $p\gamma$ (as mentioned in Ref. \cite{vogel2017diffuse}), but there is no definitive evidence that IceCube neutrino sources are $p\gamma$ dominant. 

For the model of the extragalactic background light (EBL) we choose the ”Best-fit” model from Ref.~\cite{Kneiske:2003tx} (while we have verified that the other EBL models, such as the Inoue Lower-Pop-III Model\cite{Inoue_2013} and the Stecker (upper distribution)\cite{Stecker:2005qs}, lead to only quite marginal variations in spectra).

\subsection{Potential mixing at sources: axion injection spectrum}
\label{subsec:axinjection}

If strong mixing is present, the ratio of photons to axions is 2:1 (or equipartition among two photon polarizations and axions is established, see Ref.~\cite{Tavecchio:2014yoa}, though it can be directly seen from Eq.~\ref{eq:Transport}). 

For our simulations we consider two opposite cases: when strong mixing with that relation is present and when mixing is completely absent, in other words, when only photons are injected.


At the same time, realistic mixing at the sources is quite complex (see Ref.~\cite{Tavecchio:2014yoa}, discussed also in Ref.~\cite{Harris_2014}), but in our work we focus on the response of the IGMF-mixing to the two mentioned above different scenarios of ALP-photon injection.

\section{Results of the simulations, interpretation}
\label{sec:Results}

We proceed to the discussions of the effects and their combinations caused by ALPh mixing in the IGMF.

\subsection{Multimessenger connection}
\label{subsec:messenger}

Here in Fig.~\ref{fig:multimessenger} we illustrate fluxes of neutrinos (the single power law taken as in Ref.~\cite{naab2023measurement}, shown per flavor on the graph), and its photon/axion counterpart. Presented are the cases of the "strong mixing" at the sources and "no mixing" at the sources (mixing in the IGMF happens both at IceCube and Fermi energies in both cases). For comparison and demonstrative purposes the isotropic gamma-ray background (IGRB) Model A from Fermi-LAT, Ref.~\cite{Ackermann_2015} with error bars is also given (see Sec.~\ref{sec:concl} for the discussions). For comparison the "standard scenario", not involving axions at all, is presented.

As far as we can see, photon to axion ratio at Fermi energies is canonical 2:1. As we also can see, the mixing in the IGMF combined with ICS and PP effectively pumps out injected axion component from most part of IceCube energies (the corresponding effect for the Fermi spectra will be discussed below in  Subec.~\ref{subsubsec:case2}, though leaving almost unchanged the highest energetic part of axion spectrum (an interesting consequence of a higher attenuation at these energies).

\subsection{Contributions to the full flux from lower redshifts} 
\label{subsec:evolv}

Here we consider axion-photon spectra from sources distributed up to certain maximal redshifts and compare them to the full flux. $z_{max}=0.1$, $z_{max}=1.0$, besides the full spectra, are taken, equivalent to a sharp cutoff in the source redshift evolution(Sec.~\ref{subsec:injection}). The mass of axion is adjusted so that it results in the cutoff of axion spectra. 

As can be seen in Fig.~\ref{fig:evolution}, "axion cutoff" region migrates leftwards to lower energies. That may be connected, first, to the evolution of the magnetic field taken (see  Subsec.~\ref{subsec:IGMF}) and, secondly, to the redshifting of the incoming photons.

Furthermore, as could be predicted, the fluxes of axions and photons reach canonical 2:1 for its parts coming from higher redshift, as propagation through distances that are large enough enables them to mix effectively.

\begin{figure}
\centering
\includegraphics[width=\linewidth]{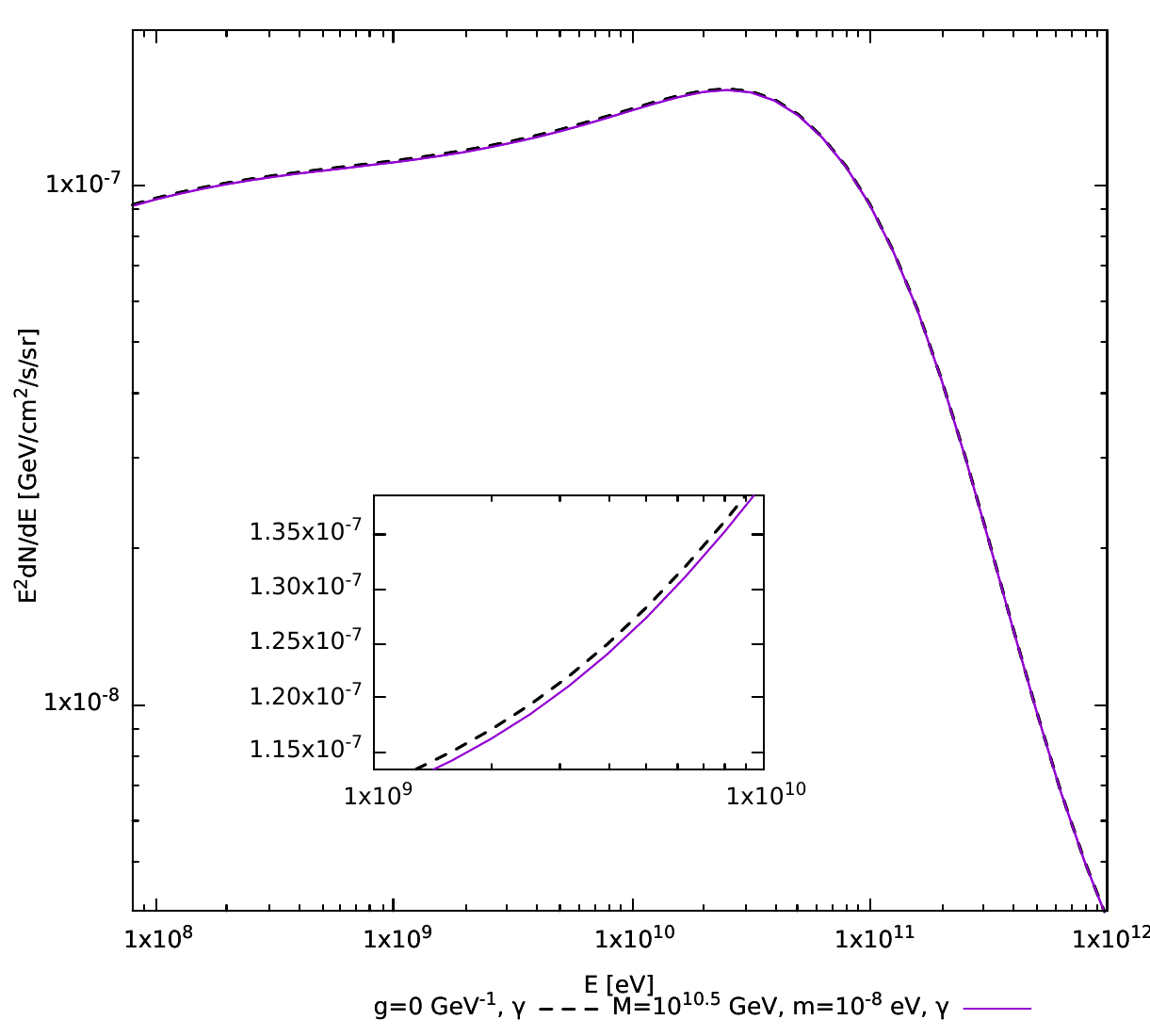}
\caption{\label{fig:30to55}
To \fullref{subsubsec:case1}.
}
\end{figure}

\begin{figure}
\centering
\includegraphics[width=\linewidth]{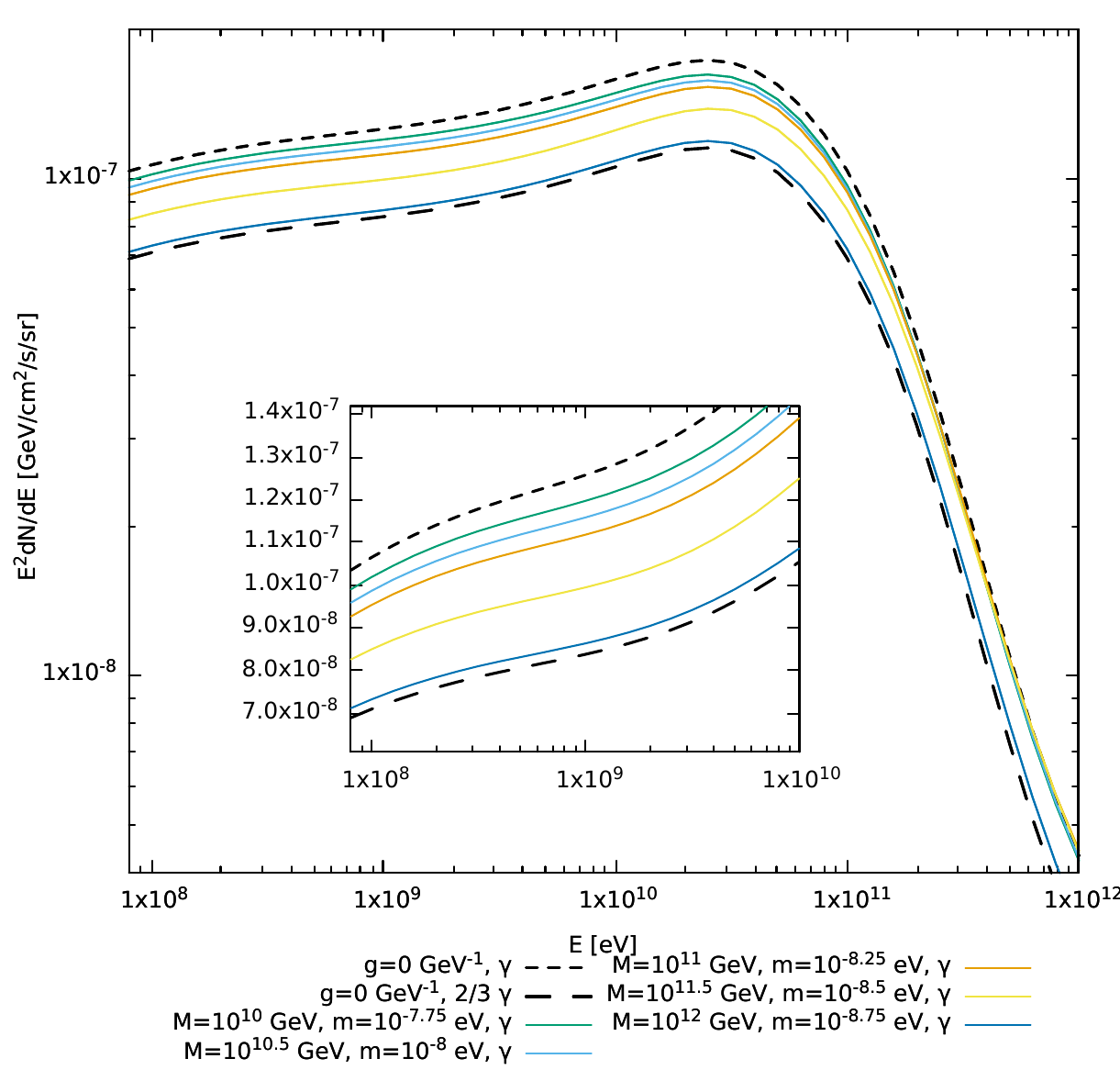}
\caption{\label{fig:axmixhigheren}
To \fullref{subsubsec:case2}; "$g=0 ~ \mbox{GeV}^{-1}, \gamma$" corresponds to the standard scenario, "$g=0 ~ \mbox{GeV}^{-1}, 2/3, \gamma$" corresponds to two thirds of that flux.
}
\end{figure}


\subsection{Various cases and scenarios}
We proceed to the classification of the scenarios. Specific areas of \Cref{fig:30to55,fig:axmixhigheren,fig:fermimixnoax,fig:fermimixwithax} are zoomed in and placed as insets, for a better comparison of photon spectra in the cases of different couplings.  
\label{subsec:cases}
\subsubsection{ALPh mixing in the IGMF at IceCube energies (no mixing at the sources)}
\label{subsubsec:case1}

Since higher energies have lower attenuation lengths of photons, neither effective oscillation length of different $M$ considered is sufficient to "capture" more ALP/photons at higher energies. Though at Fermi energies it leads to photon flux being consistently below the one in the non-axion (standard) scenario, the alternation does not exceed one percent, see Fig.~\ref{fig:30to55}.

\subsubsection{ALPh mixing in the IGMF at IceCube energies (strong mixing at the sources)}
\label{subsubsec:case2}
If the mixing in the IGMF was not present, the mixing at the sources would simply lead to the whole spectra being reduced by a factor of 3/2, however mixing at those energies frees up some axions, converting them into photons and thus letting them reach Fermi range energies, see Fig.~\ref{fig:axmixhigheren}. The effect gets more noticeable with lower $M$, however, even with $M=10^{11.5}$ GeV it remains quite sufficient and is visible even with $M=10^{12}$ GeV, which is additionally shown here (and it ceases to appear with $M=10^{12.5}$ GeV).

\subsubsection{ALPh mixing in the IGMF both at Fermi and IceCube energies (no mixing at the sources)} 
\label{subsubsec:case3}
\begin{figure}
\centering
\includegraphics[width=\linewidth]{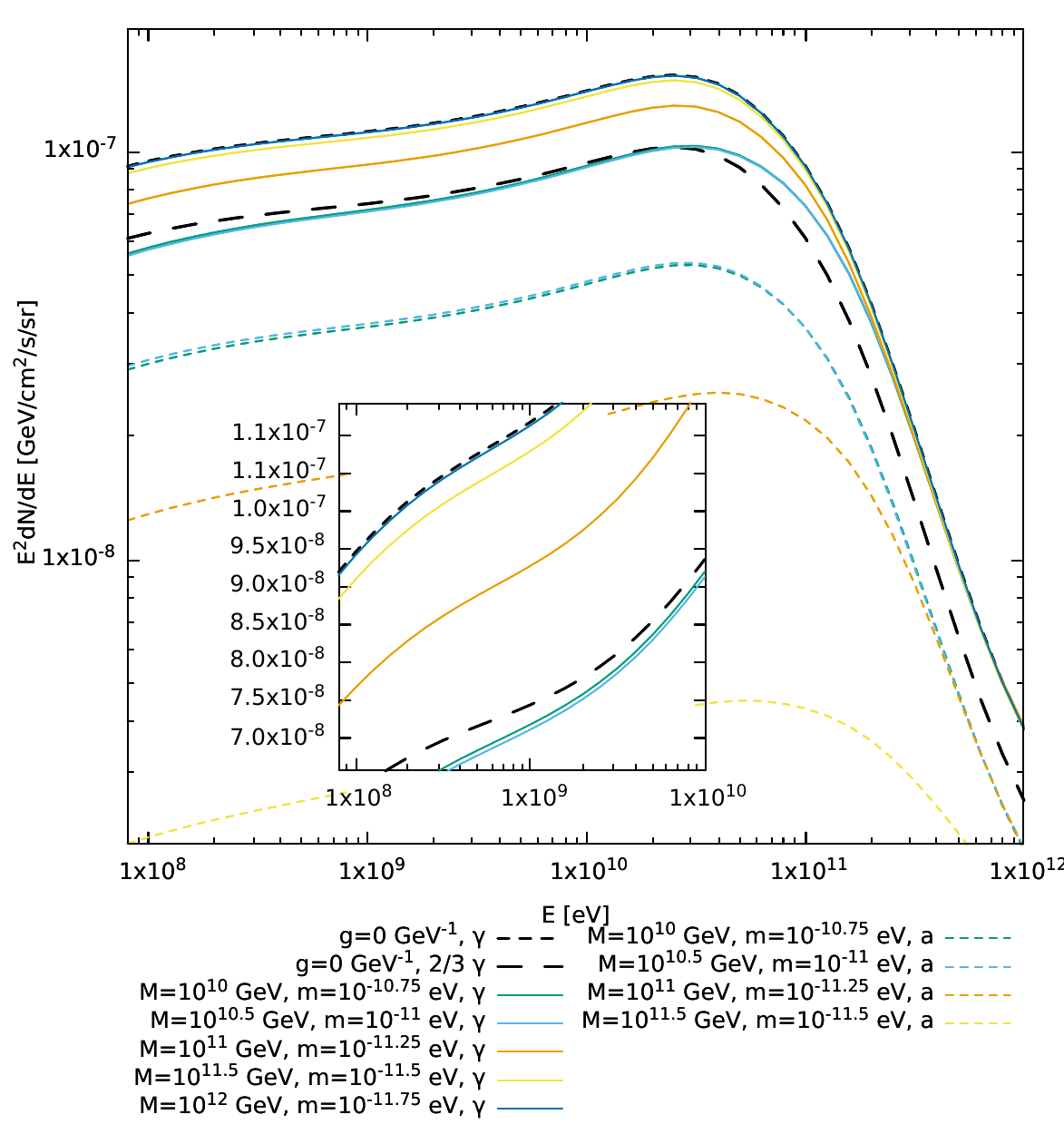}
\caption{\label{fig:fermimixnoax}
To \fullref{subsubsec:case3}; solid/dotted colored lines are used for photons/axions.
}
\end{figure}

As we can see in Fig.~\ref{fig:fermimixnoax}, the full flux is distributed as expected 2:1 for sufficiently high values of the coupling constant and deviates from that ratio only for $M=10^{11} ~\mbox{GeV}$. The flux of axions is an order of magnitude lower for the value $M=10^{11.5} ~\mbox{GeV}$, when some influence on the photon spectrum is still present.

\subsubsection{ALPh mixing in the IGMF both at Fermi and IceCube energies (strong mixing at the sources)}
\label{subsubsec:case4}
\begin{figure}
\centering
\includegraphics[width=\linewidth]{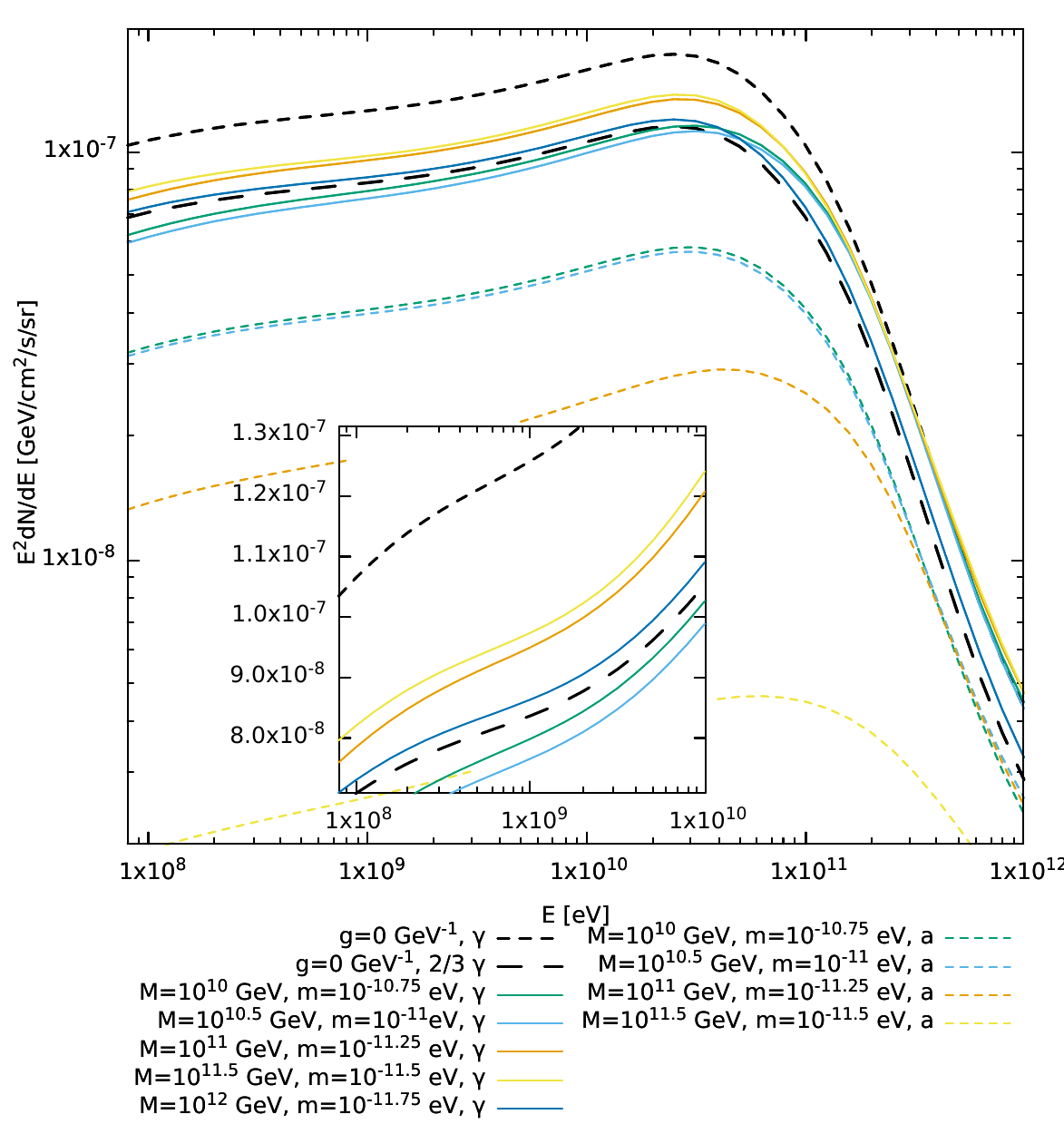}
\caption{\label{fig:fermimixwithax}
To \fullref{subsubsec:case4}; Same as in  Fig.~\ref{fig:fermimixnoax}, but for strong mixing at the sources.
}
\end{figure}

\begin{figure}
\centering
\includegraphics[width=\linewidth]{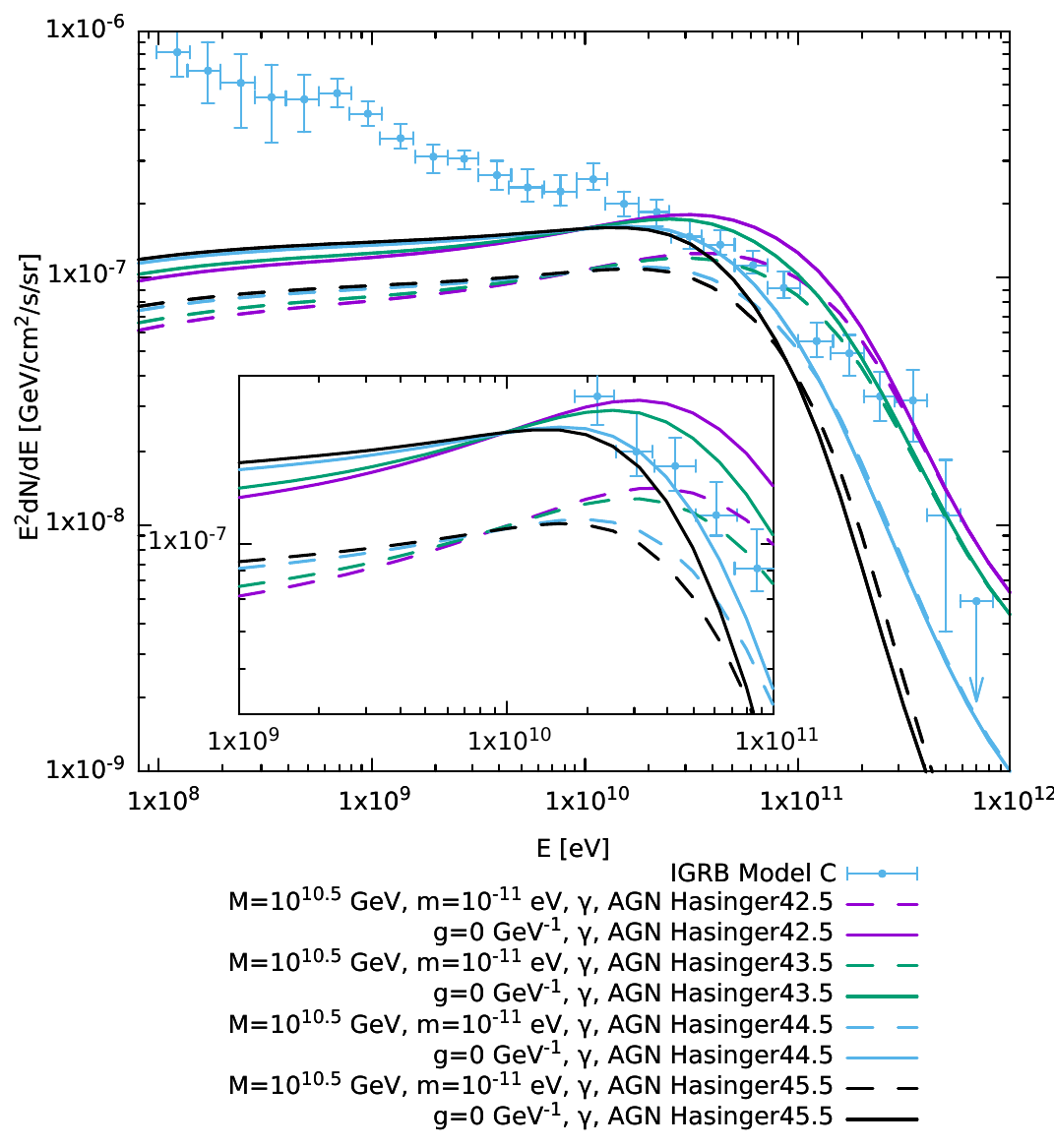}
\caption{\label{fig:comparisonredshifts}
To \fullref{subsec:injection}: outcomes for different redshift distributions, given in Ref.\cite{Hasinger:2005sb}(Table 6, plotted in Fig.9 therein); the photon flux in the standard scenario, as above, is denoted as "$g=0 ~ \mbox{GeV}^{-1}, \gamma$", and is compared to the case when ALP parameters enable conversion.
}
\end{figure}

As we can see in Fig.~\ref{fig:fermimixwithax}, the effect combines features described in Subsec.~\ref{subsubsec:case2} and ~\ref{subsubsec:case3}, namely, the mixing at Fermi energies leads to certain ratios between photons and axions at Fermi energies, and injection of both photons and axions at IceCube energies leads to a "displacement", lowering of these spectra as we described. Thus, these two mechanisms lead to similar effects, and different spectra "whirl around" two thirds of the standard scenario spectrum.


\subsection{Potential effects of later conversion in the Milky Way magnetic field}
\label{subsec:MW}

The solutions for the flux after propagation through the MW magnetic field can be found in the formalism of the same transport equations, this time not taking into account ICS and PP, passing the components $\rho_0$, $\rho_1$, $\rho_2$, $\rho_3$, $\rho_4$, $\rho_5$, $\rho_6$, $\rho_7$, $\rho_8$, $N_a$, $N_{ph1}$, $N_{ph2}$ from the output of the IGMF-solver to the input of the "MW-solver".

However, the geometry of magnetic fields in the MW is quite vague, and different models of its regular component exist (see Ref.~\cite{Simet_2008}, comparing conversion probabilities isocountours for different MW field models, thus showcasing model-dependence of that problem).

The effect of mixing itself will at best reduce the photon flux up to 2/3 of it. Due to the MW magnetic field anisotropy, from the most general considerations, the photon flux may acquire some anisotropy, which is a separate problem for discussion.  

Taking for an estimate $L \sim 10$ kpc, $B \sim 1$ mG, referring to the condition (\ref{eq:Lengthmix}),  $10^{10.5}$ GeV (close parameter see in the map of isocontours of probabilities in Ref.~\cite{Simet_2008}) corresponds to a kind of a critical, transitional case between the strong mixing with $M=10^{9.5}$ GeV and almost no mixing with $M=10^{11.5}$ GeV. From simplified logic, in the cases when the mixing in the IGMF is strong or leans towards the strong one, for $M=10^{10}$ GeV or $M=10^{10.5}$ GeV and likely for $M=10^{11}$ GeV the mixing in the IGMF is supposed to outweigh the MW and shield its effects (note that if it was not for the IGMF-mixing, there would be a possibility for the photon flux to be reduced up to two thirds of it at energies from $10^{11}$ eV to $10^{12}$ eV (see Fig.~\ref{fig:multimessenger}).

In order not to obscure the physical interpretation presented in the classification of the scenarios in Sec.~\ref{sec:Results}, here we conclude our analysis. Note that the results of the simulations are readjustable: if the magnitude of the IGMF is weaker, the considered parameters shift towards stronger coupling, and in that case the results of the paper (Sec.~\ref{sec:Results}) remain valid (but, if the MW conversion was included, numerous combinations of the MW models and the IGMF magnitudes would bring intertwined results).

\section{Discussions and conclusion}
\label{sec:concl}

Unlike in most research papers on the subject of ALPh conversion effect, in our configuration the conversion goes in parallel with the interaction of the flux with background radiation and the energies of injection and detection of the effect do not overlap. The features and possible resulting regimes are briefly described and demonstrated.

The effect discussed in the paper may alleviate the tension between neutrino and photon spectra (discussed in  Ref.~\cite{Murase_2016} or in Ref.~\cite{kistler2015problems, Kalashev_2019}), see Fig.~\ref{fig:multimessenger}, at energies from $10^{10}  ~\mbox{eV}$ to $10^{11}  ~\mbox{eV}$, however it leaves the flux of photons unchanged at energies from to $10^{11}  ~\mbox{eV}$ to $10^{12}  ~\mbox{eV}$.

\section*{Acknowledgements}

The author is indebted to S.~Troitsky for the original idea of the manuscript and valuable remarks throughout the entire work and much obliged to O.~Kalashev, the creator of the tool TransportCR, for the helpful directions regarding its use and discussions devoted to the modification of its code.

This work is supported by the RSF grant 22-12-00253. Numerical calculations were performed on the Computational Cluster of the Theoretical division of INR RAS. The author thanks the Theoretical Physics and Mathematics Advancement Foundation “BASIS” for the fellowship under the contract 22-2-1-122-1.

 \bibliographystyle{elsarticle-num} 
 \bibliography{flav1}

\begin{thebibliography}{10}
\expandafter\ifx\csname url\endcsname\relax
  \def\url#1{\texttt{#1}}\fi
\expandafter\ifx\csname urlprefix\endcsname\relax\def\urlprefix{URL }\fi
\expandafter\ifx\csname href\endcsname\relax
  \def\href#1#2{#2} \def\path#1{#1}\fi

\bibitem{Mastrototaro_2022}
L.~Mastrototaro, P.~Carenza, M.~Chianese, D.~F.~G. Fiorillo, G.~Miele, A.~Mirizzi, D.~Montanino, {Constraining axion-like particles with the diffuse gamma-ray flux measured by the Large High Altitude Air Shower Observatory}, Eur. Phys. J. C 82~(11) (2022) 1012.
\newblock \href {http://arxiv.org/abs/2206.08945} {\path{arXiv:2206.08945}}, \href {https://doi.org/10.1140/epjc/s10052-022-10979-6} {\path{doi:10.1140/epjc/s10052-022-10979-6}}.

\bibitem{Troitsky_2022}
S.~V. Troitsky, \href{https://doi.org/10.1134%2Fs0021364022602408}{{Parameters of Axion-Like Particles Required to Explain High-Energy Photons from {GRB} 221009A}}, {JETP} Letters 116~(11) (2022) 767--770.
\newblock \href {https://doi.org/10.1134/s0021364022602408} {\path{doi:10.1134/s0021364022602408}}.
\newline\urlprefix\url{https://doi.org/10.1134%2Fs0021364022602408}

\bibitem{Ajello_2016}
M.~Ajello, et~al., {Search for Spectral Irregularities due to Photon\textendash{}Axionlike-Particle Oscillations with the Fermi Large Area Telescope}, Phys. Rev. Lett. 116~(16) (2016) 161101.
\newblock \href {http://arxiv.org/abs/1603.06978} {\path{arXiv:1603.06978}}, \href {https://doi.org/10.1103/PhysRevLett.116.161101} {\path{doi:10.1103/PhysRevLett.116.161101}}.

\bibitem{Galanti:2022iwb}
G.~Galanti, {Photon-ALP interaction as a measure of initial photon polarization}, Phys. Rev. D 105~(8) (2022) 083022.
\newblock \href {http://arxiv.org/abs/2202.10315} {\path{arXiv:2202.10315}}, \href {https://doi.org/10.1103/PhysRevD.105.083022} {\path{doi:10.1103/PhysRevD.105.083022}}.

\bibitem{DeAngelis:2011id}
A.~De~Angelis, G.~Galanti, M.~Roncadelli, {Relevance of axion-like particles for very-high-energy astrophysics}, Phys. Rev. D 84 (2011) 105030, [Erratum: Phys.Rev.D 87, 109903 (2013)].
\newblock \href {http://arxiv.org/abs/1106.1132} {\path{arXiv:1106.1132}}, \href {https://doi.org/10.1103/PhysRevD.84.105030} {\path{doi:10.1103/PhysRevD.84.105030}}.

\bibitem{Marsh_2017}
M.~D. Marsh, H.~R. Russell, A.~C. Fabian, B.~R. McNamara, P.~Nulsen, C.~S. Reynolds, \href{https://doi.org/10.1088%2F1475-7516%2F2017%2F12%2F036}{A new bound on axion-like particles}, Journal of Cosmology and Astroparticle Physics 2017~(12) (2017) 036--036.
\newblock \href {https://doi.org/10.1088/1475-7516/2017/12/036} {\path{doi:10.1088/1475-7516/2017/12/036}}.
\newline\urlprefix\url{https://doi.org/10.1088%2F1475-7516%2F2017%2F12%2F036}

\bibitem{Liang_2021}
Y.-F. Liang, X.-F. Zhang, J.-G. Cheng, H.-D. Zeng, Y.-Z. Fan, E.-W. Liang, \href{https://doi.org/10.1088%2F1475-7516%2F2021%2F11%2F030}{Effect of axion-like particles on the spectrum of the extragalactic gamma-ray background}, Journal of Cosmology and Astroparticle Physics 2021~(11) (2021) 030.
\newblock \href {https://doi.org/10.1088/1475-7516/2021/11/030} {\path{doi:10.1088/1475-7516/2021/11/030}}.
\newline\urlprefix\url{https://doi.org/10.1088%2F1475-7516%2F2021%2F11%2F030}

\bibitem{Irastorza_2018}
I.~G. Irastorza, J.~Redondo, \href{https://doi.org/10.1016%2Fj.ppnp.2018.05.003}{New experimental approaches in the search for axion-like particles}, Progress in Particle and Nuclear Physics 102 (2018) 89--159.
\newblock \href {https://doi.org/10.1016/j.ppnp.2018.05.003} {\path{doi:10.1016/j.ppnp.2018.05.003}}.
\newline\urlprefix\url{https://doi.org/10.1016%2Fj.ppnp.2018.05.003}

\bibitem{Guo_2021}
J.~Guo, H.-J. Li, X.-J. Bi, S.-J. Lin, P.-F. Yin, {Implications of axion-like particles from the Fermi-LAT and H.E.S.S. observations of PG 1553+113 and PKS 2155\ensuremath{-}304}, Chin. Phys. C 45~(2) (2021) 025105.
\newblock \href {http://arxiv.org/abs/2002.07571} {\path{arXiv:2002.07571}}, \href {https://doi.org/10.1088/1674-1137/abcd2e} {\path{doi:10.1088/1674-1137/abcd2e}}.

\bibitem{meyer2016searches}
M.~Meyer, Searches for axionlike particles using gamma-ray observations (2016).
\newblock \href {http://arxiv.org/abs/1611.07784} {\path{arXiv:1611.07784}}.

\bibitem{Troitsky:2023uwu}
S.~Troitsky, {Towards a model of photon-axion conversion in the host galaxy of GRB 221009A} (7 2023).
\newblock \href {http://arxiv.org/abs/2307.08313} {\path{arXiv:2307.08313}}.

\bibitem{Pant:2023khq}
B.~P. Pant, Sunanda, R.~Moharana, S.~S., {Implications of photon-ALP oscillations in the extragalactic neutrino source TXS 0506+056 at sub-PeV energies}, Phys. Rev. D 108~(2) (2023) 023016.
\newblock \href {https://doi.org/10.1103/PhysRevD.108.023016} {\path{doi:10.1103/PhysRevD.108.023016}}.

\bibitem{De_Angelis_2007}
A.~D. Angelis, M.~Roncadelli, O.~Mansutti, \href{https://doi.org/10.1103%2Fphysrevd.76.121301}{Evidence for a new light spin-zero boson from cosmological gamma-ray propagation?}, Physical Review D 76~(12) (dec 2007).
\newblock \href {https://doi.org/10.1103/physrevd.76.121301} {\path{doi:10.1103/physrevd.76.121301}}.
\newline\urlprefix\url{https://doi.org/10.1103%2Fphysrevd.76.121301}

\bibitem{Mirizzi_2017}
A.~Mirizzi, D.~Montanino, \href{https://doi.org/10.1088%2F1475-7516%2F2009%2F12%2F004}{Stochastic conversions of {TeV} photons into axion-like particles in extragalactic magnetic fields}, Journal of Cosmology and Astroparticle Physics 2009~(12) (2017) 004--004.
\newblock \href {https://doi.org/10.1088/1475-7516/2009/12/004} {\path{doi:10.1088/1475-7516/2009/12/004}}.
\newline\urlprefix\url{https://doi.org/10.1088%2F1475-7516%2F2009%2F12%2F004}

\bibitem{Galanti:2018myb}
G.~Galanti, M.~Roncadelli, {Extragalactic photon\textendash{}axion-like particle oscillations up to 1000 TeV}, JHEAp 20 (2018) 1--17.
\newblock \href {http://arxiv.org/abs/1805.12055} {\path{arXiv:1805.12055}}, \href {https://doi.org/10.1016/j.jheap.2018.07.002} {\path{doi:10.1016/j.jheap.2018.07.002}}.

\bibitem{S_nchez_Conde_2009}
M.~A. S{\'{a} }nchez-Conde, D.~Paneque, E.~Bloom, F.~Prada, A.~Dom{\'{\i}}nguez, \href{https://doi.org/10.1103%2Fphysrevd.79.123511}{Hints of the existence of axionlike particles from the gamma-ray spectra of cosmological sources}, Physical Review D 79~(12) (jun 2009).
\newblock \href {https://doi.org/10.1103/physrevd.79.123511} {\path{doi:10.1103/physrevd.79.123511}}.
\newline\urlprefix\url{https://doi.org/10.1103%2Fphysrevd.79.123511}

\bibitem{Kachelriess:2023fta}
M.~Kachelriess, J.~Tjemsland, {Detecting ALP wiggles at TeV energies} (5 2023).
\newblock \href {http://arxiv.org/abs/2305.03604} {\path{arXiv:2305.03604}}.

\bibitem{Cs_ki_2002}
C.~Cs{\'{a} }ki, N.~Kaloper, J.~Terning, \href{https://doi.org/10.1103%2Fphysrevlett.88.161302}{Dimming supernovae without cosmic acceleration}, Physical Review Letters 88~(16) (apr 2002).
\newblock \href {https://doi.org/10.1103/physrevlett.88.161302} {\path{doi:10.1103/physrevlett.88.161302}}.
\newline\urlprefix\url{https://doi.org/10.1103%2Fphysrevlett.88.161302}

\bibitem{M_rtsell_2002}
E.~Mortsell, L.~Bergstrom, A.~Goobar, {Photon axion oscillations and type Ia supernovae}, Phys. Rev. D 66 (2002) 047702.
\newblock \href {http://arxiv.org/abs/astro-ph/0202153} {\path{arXiv:astro-ph/0202153}}, \href {https://doi.org/10.1103/PhysRevD.66.047702} {\path{doi:10.1103/PhysRevD.66.047702}}.

\bibitem{Christensson_2003}
M.~Christensson, M.~Fairbairn, \href{https://doi.org/10.1016%2Fs0370-2693%2803%2900641-5}{Photon{\textendash}axion mixing in an inhomogeneous universe}, Physics Letters B 565 (2003) 10--18.
\newblock \href {https://doi.org/10.1016/s0370-2693(03)00641-5} {\path{doi:10.1016/s0370-2693(03)00641-5}}.
\newline\urlprefix\url{https://doi.org/10.1016%2Fs0370-2693%2803%2900641-5}

\bibitem{Kachelriess:2023cjh}
M.~Kachelriess, J.~Tjemsland, {Photon-ALP oscillations at CTA energies}, SciPost Phys. Proc. 12 (2023) 043.
\newblock \href {https://doi.org/10.21468/SciPostPhysProc.12.043} {\path{doi:10.21468/SciPostPhysProc.12.043}}.

\bibitem{Galanti_2018}
G.~Galanti, M.~Roncadelli, \href{https://doi.org/10.1103%2Fphysrevd.98.043018}{Behavior of axionlike particles in smoothed out domainlike magnetic fields}, Physical Review D 98~(4) (aug 2018).
\newblock \href {https://doi.org/10.1103/physrevd.98.043018} {\path{doi:10.1103/physrevd.98.043018}}.
\newline\urlprefix\url{https://doi.org/10.1103%2Fphysrevd.98.043018}

\bibitem{Meyer_2021}
M.~Meyer, J.~Davies, J.~Kuhlmann, \href{https://doi.org/10.22323%2F1.395.0557}{{gammaALPs}: An open-source python package for computing photon-axion-like-particle oscillations in astrophysical environments}, in: Proceedings of 37th International Cosmic Ray Conference {\textemdash} {PoS}({ICRC}2021), Sissa Medialab, 2021.
\newblock \href {https://doi.org/10.22323/1.395.0557} {\path{doi:10.22323/1.395.0557}}.
\newline\urlprefix\url{https://doi.org/10.22323%2F1.395.0557}

\bibitem{Kachelriess:2021rzc}
M.~Kachelriess, J.~Tjemsland, {On the origin and the detection of characteristic axion wiggles in photon spectra}, JCAP 01~(01) (2022) 025.
\newblock \href {http://arxiv.org/abs/2111.08303} {\path{arXiv:2111.08303}}, \href {https://doi.org/10.1088/1475-7516/2022/01/025} {\path{doi:10.1088/1475-7516/2022/01/025}}.

\bibitem{Kalashev_2019}
O.~E. Kalashev, A.~Kusenko, E.~Vitagliano, {Cosmic infrared background excess from axionlike particles and implications for multimessenger observations of blazars}, Phys. Rev. D 99~(2) (2019) 023002.
\newblock \href {http://arxiv.org/abs/1808.05613} {\path{arXiv:1808.05613}}, \href {https://doi.org/10.1103/PhysRevD.99.023002} {\path{doi:10.1103/PhysRevD.99.023002}}.

\bibitem{Berezinsky_2016}
V.~Berezinsky, O.~Kalashev, \href{https://doi.org/10.1103%2Fphysrevd.94.023007}{High-energy electromagnetic cascades in extragalactic space: Physics and features}, Physical Review D 94~(2) (jul 2016).
\newblock \href {https://doi.org/10.1103/physrevd.94.023007} {\path{doi:10.1103/physrevd.94.023007}}.
\newline\urlprefix\url{https://doi.org/10.1103%2Fphysrevd.94.023007}

\bibitem{Blanco_2019}
C.~Blanco, \href{https://doi.org/10.1088%2F1475-7516%2F2019%2F01%2F013}{$\gamma$-cascade: a simple program to compute cosmological gamma-ray propagation}, Journal of Cosmology and Astroparticle Physics 2019~(01) (2019) 013--013.
\newblock \href {https://doi.org/10.1088/1475-7516/2019/01/013} {\path{doi:10.1088/1475-7516/2019/01/013}}.
\newline\urlprefix\url{https://doi.org/10.1088%2F1475-7516%2F2019%2F01%2F013}

\bibitem{Batista_2016}
R.~A. Batista, A.~Dundovic, M.~Erdmann, K.-H. Kampert, D.~Kuempel, G.~Müller, G.~Sigl, A.~van Vliet, D.~Walz, T.~Winchen, \href{https://doi.org/10.1088%2F1475-7516%2F2016%2F05%2F038}{{CRPropa} 3{\textemdash}a public astrophysical simulation framework for propagating extraterrestrial ultra-high energy particles}, Journal of Cosmology and Astroparticle Physics 2016~(05) (2016) 038--038.
\newblock \href {https://doi.org/10.1088/1475-7516/2016/05/038} {\path{doi:10.1088/1475-7516/2016/05/038}}.
\newline\urlprefix\url{https://doi.org/10.1088%2F1475-7516%2F2016%2F05%2F038}

\bibitem{Palladino_2020}
A.~Palladino, M.~Spurio, F.~Vissani, \href{https://doi.org/10.3390%2Funiverse6020030}{Neutrino telescopes and high-energy cosmic neutrinos}, Universe 6~(2) (2020) 30.
\newblock \href {https://doi.org/10.3390/universe6020030} {\path{doi:10.3390/universe6020030}}.
\newline\urlprefix\url{https://doi.org/10.3390%2Funiverse6020030}

\bibitem{Fornasa_2015}
M.~Fornasa, M.~A. S{\'{a}}nchez-Conde, \href{https://doi.org/10.1016%2Fj.physrep.2015.09.002}{The nature of the diffuse gamma-ray background}, Physics Reports 598 (2015) 1--58.
\newblock \href {https://doi.org/10.1016/j.physrep.2015.09.002} {\path{doi:10.1016/j.physrep.2015.09.002}}.
\newline\urlprefix\url{https://doi.org/10.1016%2Fj.physrep.2015.09.002}

\bibitem{Eckner_2022}
C.~Eckner, F.~Calore, {First constraints on axionlike particles from Galactic sub-PeV gamma rays}, Phys. Rev. D 106~(8) (2022) 083020.
\newblock \href {http://arxiv.org/abs/2204.12487} {\path{arXiv:2204.12487}}, \href {https://doi.org/10.1103/PhysRevD.106.083020} {\path{doi:10.1103/PhysRevD.106.083020}}.

\bibitem{Murase_2016}
K.~Murase, D.~Guetta, M.~Ahlers, \href{https://doi.org/10.1103%2Fphysrevlett.116.071101}{Hidden cosmic-ray accelerators as an origin of {TeV}-{PeV} cosmic neutrinos}, Physical Review Letters 116~(7) (feb 2016).
\newblock \href {https://doi.org/10.1103/physrevlett.116.071101} {\path{doi:10.1103/physrevlett.116.071101}}.
\newline\urlprefix\url{https://doi.org/10.1103%2Fphysrevlett.116.071101}

\bibitem{Kovalev_2022}
Y.~Y. Kovalev, A.~V. Plavin, S.~V. Troitsky, \href{https://doi.org/10.3847%2F2041-8213%2Faca1ae}{Galactic contribution to the high-energy neutrino flux found in track-like {IceCube} events}, The Astrophysical Journal Letters 940~(2) (2022) L41.
\newblock \href {https://doi.org/10.3847/2041-8213/aca1ae} {\path{doi:10.3847/2041-8213/aca1ae}}.
\newline\urlprefix\url{https://doi.org/10.3847%2F2041-8213%2Faca1ae}

\bibitem{Albert_2023}
A.~Albert, et~al., {Hint for a TeV neutrino emission from the Galactic Ridge with ANTARES}, Phys. Lett. B 841 (2023) 137951.
\newblock \href {http://arxiv.org/abs/2212.11876} {\path{arXiv:2212.11876}}, \href {https://doi.org/10.1016/j.physletb.2023.137951} {\path{doi:10.1016/j.physletb.2023.137951}}.

\bibitem{2023}
R.~Abbasi, et~al., {Observation of high-energy neutrinos from the Galactic plane}, Science 380~(6652) (2023) adc9818.
\newblock \href {http://arxiv.org/abs/2307.04427} {\path{arXiv:2307.04427}}, \href {https://doi.org/10.1126/science.adc9818} {\path{doi:10.1126/science.adc9818}}.

\bibitem{IceCube:2020wum}
R.~Abbasi, et~al., {The IceCube high-energy starting event sample: Description and flux characterization with 7.5 years of data}, Phys. Rev. D 104 (2021) 022002.
\newblock \href {http://arxiv.org/abs/2011.03545} {\path{arXiv:2011.03545}}, \href {https://doi.org/10.1103/PhysRevD.104.022002} {\path{doi:10.1103/PhysRevD.104.022002}}.

\bibitem{kistler2015problems}
M.~D. Kistler, {Problems and Prospects from a Flood of Extragalactic TeV Neutrinos in IceCube} (11 2015).
\newblock \href {http://arxiv.org/abs/1511.01530} {\path{arXiv:1511.01530}}.

\bibitem{Kun:2023uld}
E.~Kun, I.~Bartos, J.~Becker~Tjus, P.~L. Biermann, A.~Franckowiak, F.~Halzen, G.~Mezo, {Searching for temporary gamma-ray dark blazars associated with IceCube neutrinos}, Astron. Astrophys. 679 (2023) A46.
\newblock \href {http://arxiv.org/abs/2305.06729} {\path{arXiv:2305.06729}}, \href {https://doi.org/10.1051/0004-6361/202346710} {\path{doi:10.1051/0004-6361/202346710}}.

\bibitem{kalashev2014simulations}
O.~E. Kalashev, E.~Kido, Simulations of ultra high energy cosmic rays propagation (2014).
\newblock \href {http://arxiv.org/abs/1406.0735} {\path{arXiv:1406.0735}}.

\bibitem{Fairbairn_2011}
M.~Fairbairn, T.~Rashba, S.~Troitsky, \href{https://doi.org/10.1103%2Fphysrevd.84.125019}{{Photon-axion mixing and ultra-high energy cosmic rays from {BL} Lac type objects: Shining light through the Universe}}, Physical Review D 84~(12) (dec 2011).
\newblock \href {https://doi.org/10.1103/physrevd.84.125019} {\path{doi:10.1103/physrevd.84.125019}}.
\newline\urlprefix\url{https://doi.org/10.1103%2Fphysrevd.84.125019}

\bibitem{vogel2017diffuse}
H.~Vogel, R.~Laha, M.~Meyer, Diffuse axion-like particle searches (2017).
\newblock \href {http://arxiv.org/abs/1712.01839} {\path{arXiv:1712.01839}}.

\bibitem{Kartavtsev_2017}
A.~Kartavtsev, G.~Raffelt, H.~Vogel, \href{https://doi.org/10.1088%2F1475-7516%2F2017%2F01%2F024}{Extragalactic photon-{ALP} conversion at {CTA} energies}, Journal of Cosmology and Astroparticle Physics 2017~(01) (2017) 024--024.
\newblock \href {https://doi.org/10.1088/1475-7516/2017/01/024} {\path{doi:10.1088/1475-7516/2017/01/024}}.
\newline\urlprefix\url{https://doi.org/10.1088%2F1475-7516%2F2017%2F01%2F024}

\bibitem{Zhang_2013}
Y.~Zhang, A.~Burrows, {Transport Equations for Oscillating Neutrinos}, Phys. Rev. D 88~(10) (2013) 105009.
\newblock \href {http://arxiv.org/abs/1310.2164} {\path{arXiv:1310.2164}}, \href {https://doi.org/10.1103/PhysRevD.88.105009} {\path{doi:10.1103/PhysRevD.88.105009}}.

\bibitem{GellRef2}
Properties of the gell-mann matrices, \url{https://scipp.ucsc.edu/~haber/archives/physics251_17/gellmann17.pdf}, accessed: 2024-10-25.

\bibitem{Lee_1998}
S.~Lee, \href{https://doi.org/10.1103%2Fphysrevd.58.043004}{Propagation of extragalactic high energy cosmic and $\gamma$ rays}, Physical Review D 58~(4) (jul 1998).
\newblock \href {https://doi.org/10.1103/physrevd.58.043004} {\path{doi:10.1103/physrevd.58.043004}}.
\newline\urlprefix\url{https://doi.org/10.1103%2Fphysrevd.58.043004}

\bibitem{ParticleDataGroup:2022pth}
R.~L. Workman, et~al., {Review of Particle Physics}, PTEP 2022 (2022) 083C01.
\newblock \href {https://doi.org/10.1093/ptep/ptac097} {\path{doi:10.1093/ptep/ptac097}}.

\bibitem{Meyer_2017}
M.~Meyer, D.~Montanino, J.~Conrad, \href{https://doi.org/10.1088%2F1475-7516%2F2014%2F09%2F003}{On detecting oscillations of gamma rays into axion-like particles in turbulent and coherent magnetic fields}, Journal of Cosmology and Astroparticle Physics 2014~(09) (2017) 003--003.
\newblock \href {https://doi.org/10.1088/1475-7516/2014/09/003} {\path{doi:10.1088/1475-7516/2014/09/003}}.
\newline\urlprefix\url{https://doi.org/10.1088%2F1475-7516%2F2014%2F09%2F003}

\bibitem{Blytt_2020}
M.~Blytt, M.~Kachelrie{\ss}, S.~Ostapchenko, \href{https://doi.org/10.1016%2Fj.cpc.2020.107163}{{ELMAG} 3.01: A three-dimensional monte carlo simulation of electromagnetic cascades on the extragalactic background light and in magnetic fields}, Computer Physics Communications 252 (2020) 107163.
\newblock \href {https://doi.org/10.1016/j.cpc.2020.107163} {\path{doi:10.1016/j.cpc.2020.107163}}.
\newline\urlprefix\url{https://doi.org/10.1016%2Fj.cpc.2020.107163}

\bibitem{Pshirkov_2016}
M.~S. Pshirkov, P.~G. Tinyakov, F.~R. Urban, {New limits on extragalactic magnetic fields from rotation measures}, Phys. Rev. Lett. 116~(19) (2016) 191302.
\newblock \href {http://arxiv.org/abs/1504.06546} {\path{arXiv:1504.06546}}, \href {https://doi.org/10.1103/PhysRevLett.116.191302} {\path{doi:10.1103/PhysRevLett.116.191302}}.

\bibitem{Planck:2015zrl}
P.~A.~R. Ade, et~al., {Planck 2015 results. XIX. Constraints on primordial magnetic fields}, Astron. Astrophys. 594 (2016) A19.
\newblock \href {http://arxiv.org/abs/1502.01594} {\path{arXiv:1502.01594}}, \href {https://doi.org/10.1051/0004-6361/201525821} {\path{doi:10.1051/0004-6361/201525821}}.

\bibitem{Neronov:2021xua}
A.~Neronov, D.~Semikoz, O.~Kalashev, {Limit on intergalactic magnetic field from ultra-high-energy cosmic ray hotspot in Perseus-Pisces region} (12 2021).
\newblock \href {http://arxiv.org/abs/2112.08202} {\path{arXiv:2112.08202}}.

\bibitem{Grasso_2001}
D.~Grasso, H.~R. Rubinstein, \href{https://doi.org/10.1016%2Fs0370-1573%2800%2900110-1}{{Magnetic fields in the early Universe}}, Physics Reports 348~(3) (2001) 163--266.
\newblock \href {https://doi.org/10.1016/s0370-1573(00)00110-1} {\path{doi:10.1016/s0370-1573(00)00110-1}}.
\newline\urlprefix\url{https://doi.org/10.1016%2Fs0370-1573%2800%2900110-1}

\bibitem{naab2023measurement}
R.~Naab, E.~Ganster, Z.~Zhang, {Measurement of the astrophysical diffuse neutrino flux in a combined fit of IceCube's high energy neutrino data}, in: {38th International Cosmic Ray Conference}, 2023.
\newblock \href {http://arxiv.org/abs/2308.00191} {\path{arXiv:2308.00191}}.

\bibitem{Ackermann_2015}
M.~Ackermann, et~al., {The spectrum of isotropic diffuse gamma-ray emission between 100 MeV and 820 GeV}, Astrophys. J. 799 (2015) 86.
\newblock \href {http://arxiv.org/abs/1410.3696} {\path{arXiv:1410.3696}}, \href {https://doi.org/10.1088/0004-637X/799/1/86} {\path{doi:10.1088/0004-637X/799/1/86}}.

\bibitem{Kachelrie__2017}
M.~Kachelrie{\ss}, O.~Kalashev, S.~Ostapchenko, D.~Semikoz, \href{https://doi.org/10.1103%2Fphysrevd.96.083006}{Minimal model for extragalactic cosmic rays and neutrinos}, Physical Review D 96~(8) (oct 2017).
\newblock \href {https://doi.org/10.1103/physrevd.96.083006} {\path{doi:10.1103/physrevd.96.083006}}.
\newline\urlprefix\url{https://doi.org/10.1103%2Fphysrevd.96.083006}

\bibitem{Hasinger:2005sb}
G.~Hasinger, T.~Miyaji, M.~Schmidt, {Luminosity-dependent evolution of soft x-ray selected AGN: New Chandra and XMM-Newton surveys}, Astron. Astrophys. 441 (2005) 417--434.
\newblock \href {http://arxiv.org/abs/astro-ph/0506118} {\path{arXiv:astro-ph/0506118}}, \href {https://doi.org/10.1051/0004-6361:20042134} {\path{doi:10.1051/0004-6361:20042134}}.

\bibitem{Ahlers_2018}
M.~Ahlers, F.~Halzen, \href{https://doi.org/10.1016%2Fj.ppnp.2018.05.001}{Opening a new window onto the universe with {IceCube}}, Progress in Particle and Nuclear Physics 102 (2018) 73--88.
\newblock \href {https://doi.org/10.1016/j.ppnp.2018.05.001} {\path{doi:10.1016/j.ppnp.2018.05.001}}.
\newline\urlprefix\url{https://doi.org/10.1016%2Fj.ppnp.2018.05.001}

\bibitem{Kneiske:2003tx}
T.~M. Kneiske, T.~Bretz, K.~Mannheim, D.~H. Hartmann, {Implications of cosmological gamma-ray absorption. 2. Modification of gamma-ray spectra}, Astron. Astrophys. 413 (2004) 807--815.
\newblock \href {http://arxiv.org/abs/astro-ph/0309141} {\path{arXiv:astro-ph/0309141}}, \href {https://doi.org/10.1051/0004-6361:20031542} {\path{doi:10.1051/0004-6361:20031542}}.

\bibitem{Inoue_2013}
Y.~Inoue, S.~Inoue, M.~A.~R. Kobayashi, R.~Makiya, Y.~Niino, T.~Totani, \href{http://dx.doi.org/10.1088/0004-637X/768/2/197}{Extragalactic background light from hierarchical galaxy formation: Gamma-ray attenuation up to the epoch of cosmic reionization and the first stars}, The Astrophysical Journal 768~(2) (2013) 197.
\newblock \href {https://doi.org/10.1088/0004-637x/768/2/197} {\path{doi:10.1088/0004-637x/768/2/197}}.
\newline\urlprefix\url{http://dx.doi.org/10.1088/0004-637X/768/2/197}

\bibitem{Stecker:2005qs}
F.~W. Stecker, M.~A. Malkan, S.~T. Scully, {Intergalactic photon spectra from the far ir to the uv lyman limit for 0 \ensuremath{<} Z \ensuremath{<} 6 and the optical depth of the universe to high energy gamma-rays}, Astrophys. J. 648 (2006) 774--783.
\newblock \href {http://arxiv.org/abs/astro-ph/0510449} {\path{arXiv:astro-ph/0510449}}, \href {https://doi.org/10.1086/506188} {\path{doi:10.1086/506188}}.

\bibitem{Tavecchio:2014yoa}
F.~Tavecchio, M.~Roncadelli, G.~Galanti, {Photons to axion-like particles conversion in Active Galactic Nuclei}, Phys. Lett. B 744 (2015) 375--379.
\newblock \href {http://arxiv.org/abs/1406.2303} {\path{arXiv:1406.2303}}, \href {https://doi.org/10.1016/j.physletb.2015.04.017} {\path{doi:10.1016/j.physletb.2015.04.017}}.

\bibitem{Harris_2014}
J.~Harris, P.~Chadwick, \href{https://doi.org/10.1088%2F1475-7516%2F2014%2F10%2F018}{Photon-axion mixing within the jets of active galactic nuclei and prospects for detection}, Journal of Cosmology and Astroparticle Physics 2014~(10) (2014) 018--018.
\newblock \href {https://doi.org/10.1088/1475-7516/2014/10/018} {\path{doi:10.1088/1475-7516/2014/10/018}}.
\newline\urlprefix\url{https://doi.org/10.1088%2F1475-7516%2F2014%2F10%2F018}

\bibitem{Simet_2008}
M.~Simet, D.~Hooper, P.~D. Serpico, {The Milky Way as a Kiloparsec-Scale Axionscope}, Phys. Rev. D 77 (2008) 063001.
\newblock \href {http://arxiv.org/abs/0712.2825} {\path{arXiv:0712.2825}}, \href {https://doi.org/10.1103/PhysRevD.77.063001} {\path{doi:10.1103/PhysRevD.77.063001}}.

\end{thebibliography}

\end{document}